\documentclass[trackchanges,twocolumn]
{aastex701}

\begin{document}

\title{Molecular Jets from an Evolved Protostar: Insights from JWST-ALMA Synergy}

\author[0000-0002-2338-4583]{Somnath Dutta}
\affiliation{Academia Sinica Institute of Astronomy and Astrophysics, Roosevelt Rd, Taipei 106319, Taiwan, R.O.C.}
\email{sdutta@asiaa.sinica.edu.tw}



\begin{abstract}

The combination of ALMA submillimeter and JWST/MIRI mid-infrared observations offers a transformative view of protostellar jets and outflows by probing cold and warm gas components across diverse physical conditions. We present a detailed comparison of gas distribution in these regimes for the jet/outflow system associated with G205.46$-$14.56S3 (HOPS\,315), focusing on the inner $\sim$800~AU along the jet. ALMA CO and SiO trace both lobes of the bipolar jet, revealing high-velocity collimated jets and wider outflow components. JWST/MIRI detects mainly the blueshifted lobe; the redshifted side is likely obscured by strong mid-infrared extinction. Shorter-wavelength MIRI H$_2$ rotational lines (S(7)–S(4)) trace compact jet structures resembling SiO emission, while longer-wavelength lines (S(3)–S(1)) reveal more extended emission akin to low-velocity CO.
From H$_2$ rotational emission, we identify two molecular gas temperature components: warm $(\sim 773 \pm 44~K)$ and hot $(\sim 2499\pm720~K)$. Using the ortho-to-para ratio, we estimate visual extinction $A_V \approx 23.3 \pm 2.5$~mag. JWST/MIRI emission imply a  jet mass-loss rate of $\dot{M}_{\rm J, blue, jwst} \approx (0.27 \pm 0.1) \times 10^{-6}\ M_\odot\,\mathrm{yr}^{-1}$. The combination of ALMA and JWST reveals stratified layers within the outflow and jet, as well as shock structures, providing a comprehensive view of their physical conditions. This multiwavelength study demonstrates that combining submillimeter observations from ALMA with infrared data from JWST is crucial for uncovering the full physical and chemical structure of protostellar jets and outflows.

\end{abstract}





\section{Introduction}\label{sec:intro}
Protostellar outflows and jets are key signatures of the ongoing accretion process in young stellar objects. These outflows, including both highly collimated jets and wide-angle molecular winds, are driven by the removal of excess angular momentum from the disk. This removal enables material to accrete onto the central protostar \citep{2007prpl.conf..245A,2014prpl.conf..451F,2016ARA&A..54..491B,2020A&ARv..28....1L}. 
The morphology of shocks, the presence of episodic ejections, and measurements of mass-loss rates over time all help trace the temporal behavior of accretion. 
The status of outflows and jets also informs us about the possible impact of chemical composition and planet formation in the disk \citep{2019A&A...631A..64B,2022Natur.606..272J}. 
Recent studies suggest that large dust grains grown in the inner disk are  entrained by protostellar outflows to the envelope, which eventually fall back onto the outer disk, and help circumvent the radial drift barrier—thus aiding planetesimal formation in the comet-forming region \citep[][]{2021ApJ...920L..35T}.
By observing protostellar outflows and jets — especially through multiwavelength tracers from facilities like ALMA and JWST - we can place important constraints on both the current accretion history, the chemical composition of the surrounding environment,  and the potential for planet formation in young stellar systems.

Millimeter and submillimeter observations have provided valuable insights into the structure and dynamics of protostellar outflows, revealing extended wind components, highly collimated jets, and shock-excited regions traced by molecular emission lines \citep[e.g.,][]{2014ApJ...783...29D,2015Natur.527...70P,2017NatAs...1E.152L,2021A&A...648A..45P,2021A&A...655A..65T,2024AJ....167...72D}. These wavelengths primarily trace cooler molecular gas with excitation temperatures of a few tens of Kelvin, making them especially sensitive to the entrained and swept-up material in the outflow cavity and shocked-gas along the jet axis. In contrast, the pure rotational lines of molecular hydrogen (H$_2$) at mid-infrared wavelengths serve as more direct tracers of shocked gas in relatively warmer regions within outflows \citep[e.g.,][]{1998ApJ...506L..75N,2003ApJ...590L..41L,2009ApJ...698.1244M,2010ApJ...724...69N,2011ApJ...738...80G}. Building on earlier infrared studies, recent observations with \textit{JWST}—particularly using the MIRI and NIRSpec instruments—have revealed both pure rotational and rovibrational H$_2$ lines, along with several atomic and ionized emission lines such as [Fe\,\textsc{ii}], [Ne\,\textsc{ii}], and [S\,\textsc{iii}], in protostellar outflows. These detections provide unprecedented insight into the warm and hot molecular gas components and the associated shock-excited regions in the jets and outflows \citep[e.g.,][]{2023Natur.622...48R,2024ApJ...962L..16N,2024A&A...687A..36T,2024A&A...691A.134C,2025ApJ...982..149O,2025ApJ...985..225L,2025A&A...695A.145V}.

\begin{figure}
    \centering
    \includegraphics[width=0.9\linewidth]{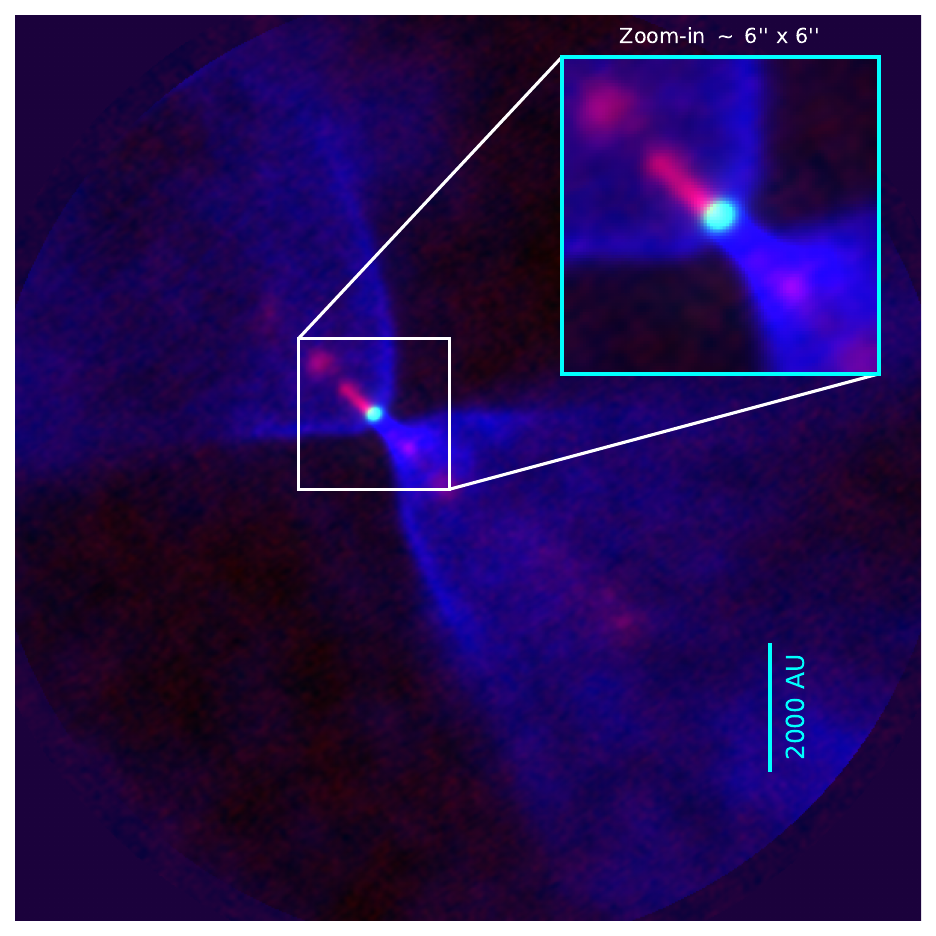}
    \caption{The three color composite image observed with ALMA ($^{12}$CO (2–1) emission at 230.53800 GHz in blue, 1.3mm Continuum in green, SiO (5–4) emission at 217.10498 GHz in red) of the G205.46$-$14.56S3 (HOPS\,315) protostellar system. The zoom-in region $\sim$ $6\arcsec \times 6\arcsec$, corresponds to the approximate coverage of JWST/MIRI Channel\ 1 for this study.}
    \label{fig:ALMA_CO_SiO_cont}
\end{figure}

G205.46$-$14.56S3 (hereafter G205S3, also known as HOPS\,315) is a Class I protostellar system characterized by a bolometric temperature of $T_{\mathrm{bol}} = 180 \pm 33$ K, bolometric luminosity $L_{\mathrm{bol}} = 6.4 \pm 2.4,L_\odot$, and a spectral index of $\sim$0.417 \citep[][]{2016ApJS..224....5F,2020ApJ...890..130T,2020ApJS..251...20D}.
\citet{2022ApJ...925...11D} detected molecular jets using ALMA observations of $^{12}$CO (2–1) line at 230.53800 GHz and SiO (5–4) line at 217.10498 GHz, reporting a moderately high jet mass-loss rate of $\dot{M}{_\mathrm{J, total, ALMA}} \sim 0.59 \times 10^{-6}\,M\odot\,\mathrm{yr}^{-1}$ and a mean jet velocity of $\sim 125\ km\ s^{-1}$. The inclination angle for G205S3 was estimated to be $40\degr \pm 8\degr$ based on the wide-angle CO outflow morphology observed with ALMA \citep[][]{2022ApJ...925...11D}. \citet[][]{2020ApJ...890..130T} performed a Gaussian fit to the dust morphology of HOPS-315, and the resulting deconvolved size corresponds to a likely inclination of $\sim47\degr$, which lies within the uncertainty range of our CO-based estimate. However, optically thick disks—common in Class 0 and early Class I sources—make it difficult to accurately constrain disk size and derived parameters from the dust continuum alone. 
Figure~\ref{fig:ALMA_CO_SiO_cont} shows the ALMA Band\,6 image of the G205S3 system, which reveals a collimated molecular jet along the axis and an extended outflow wind structure. The authors suggest that G205S3 is a relatively evolved protostar that continues to exhibit significant accretion and ejection activity, making it a compelling target for investigating outflow and jet morphology, as well as reassessing its evolutionary status through multiwavelength observations. 
Recently, \citet{2025A&A...695A.145V} have revealed that the rotational H$_2$ 0--0 S(1)--S(5) lines observed around G205S3 with \textit{JWST}/MIRI trace a warm gas component at temperatures of 500--600~K, while the ro-vibrational H$_2$ 1--0 S(0)--S(5) lines observed with \textit{JWST}/NIRSpec trace a hotter component at 2400--3800~K. They estimated an ortho-to-para ratio (OPR) of $\sim$3 from the rotational transitions and reported a decreasing OPR with distance from the inner jet to the outer regions, based on the ro-vibrational lines. This trend may be attributed to ortho-to-para conversion reactions on grain surfaces within the inner disk.

In this paper, we investigate the morphology and physical properties of the G205S3 outflow and jet by combining observations from JWST and ALMA. The complementary capabilities of JWST and ALMA enable a comprehensive characterization of the physical structure and environmental conditions of this young protostellar system. In Section 2, we describe the JWST and ALMA observations used in this study. Section 3 outlines the analysis tools developed for identifying bright spectral lines, generating moment-0 maps, and calculating rotational temperatures. In Section 4, we present a comparative analysis of the JWST and ALMA data, focusing on the physical conditions of the jet and the central protostar. Finally, Section 5 provides a summary of our main findings.


\section{Observations} \label{sec:data}
\subsection{JWST}
The JWST observations were carried out by the Mid-Infrared Instrument \citep[MIRI;][]{2015PASP..127..612B,2015PASP..127..584R,2023PASP..135d8003W}.  In this study, we utilized MIRI spectroscopy in the medium-resolution spectrometer (MRS) mode. G205S3 was observed with MIRI MRS on 2023 March 15 as part of the Cycle 1 General Observer (GO) program (PI: Melissa McClure). The spectroscopic data were taken with a standard four-point dither pattern centred on the G205S3, with coordinate at the midtime of exposure $\alpha$(J2000) = 05$^{\mathrm{h}}$46$^{\mathrm{m}}$03$\fs$63, 
$\delta$(J2000) = $-00^\circ$14$'$49$\farcs$30. The calibrated Level 3 data products were retrieved from the JWST archive. The data were processed with the JWST Calibration Software Version 1.14.0 \citep[][]{2023zndo...7577320B},  calibration reference file database version 11.17.19, and calibration reference data  $\mathrm{jwst\_1242.pmap}$.  The field of view (FOV) of the MRS increases with wavelength. The final mosaics cover areas ranging from approximately $6\arcsec \times 5\arcsec$ in Channel 1 to $10\arcsec \times 9.5\arcsec$ in Channel 4. The approximate FOV for Channel 1 is marked with a zoomed-in box in Figure~\ref{fig:ALMA_CO_SiO_cont}. The full width at half maximum (FWHM) of the point spread function (PSF) increases from short to long wavelengths from Channel 1 ($\sim 0\farcs27$)  to Channel 4 ($\sim 0\farcs96$)  \citep{2023AJ....166...45L,2023A&A...675A.111A,2023PASP..135d8003W}.

\subsection{ALMA data}
 The ALMA Band\,6 dataset for G205S3, observed on 1 November 2018, were obtained from the ALMA Science Archive \citep[Project ID: 2018.1.00302.S;][]{2020ApJS..251...20D}.
For this study, we have generated emission maps of SiO (J=5--4), CO (J=2--1), and the 1.3\,mm dust continuum. The visibility data were calibrated using the standard pipeline in CASA version 6.5.4 \citep{2007ASPC..376..127M}. All maps were produced using the \texttt{TCLEAN} task with a robust weighting factor of $+2.0$. The typical synthesized beam sizes are $\sim$0$\farcs$46\,$\times$\,0$\farcs$40 for SiO, $\sim$0$\farcs$42\,$\times$\,0$\farcs$36 for CO, and $\sim$0$\farcs$45\,$\times$\,0$\farcs$40 for the continuum. The line cubes were binned to a velocity resolution 2.0\,km\,s$^{-1}$ to improve sensitivity. The continuum peak position of G205S3 is $\alpha$(J2000) = 05$^{\mathrm{h}}$46$^{\mathrm{m}}$03$\fs$6374, 
$\delta$(J2000) = $-00^\circ$14$'$49$\farcs$603 within a primary beam width of $\sim 19\arcsec$.

\section{Analyses and Results}

\subsection{JWST: Tools for Spectral Extraction and Moment Map Generation for Emission Line Peaks}\label{sec:tools_to_detect_spectra_moment_map}
We have developed a unified tool, \texttt{peakmoment}, designed to detect bright emission lines in spectra and generate continuum-subtracted moment maps from spectral cubes. The steps are described below, and the original Python-based Jupyter notebook can be found on \texttt{zenodo}\footnote{https://doi.org/10.5281/zenodo.16623946} and \texttt{GitHub}\footnote{\texttt{peakmoment}: \url{https://github.com/somastro/peakmoment}}.


\begin{figure}
    \centering
    \includegraphics[width=1\linewidth]{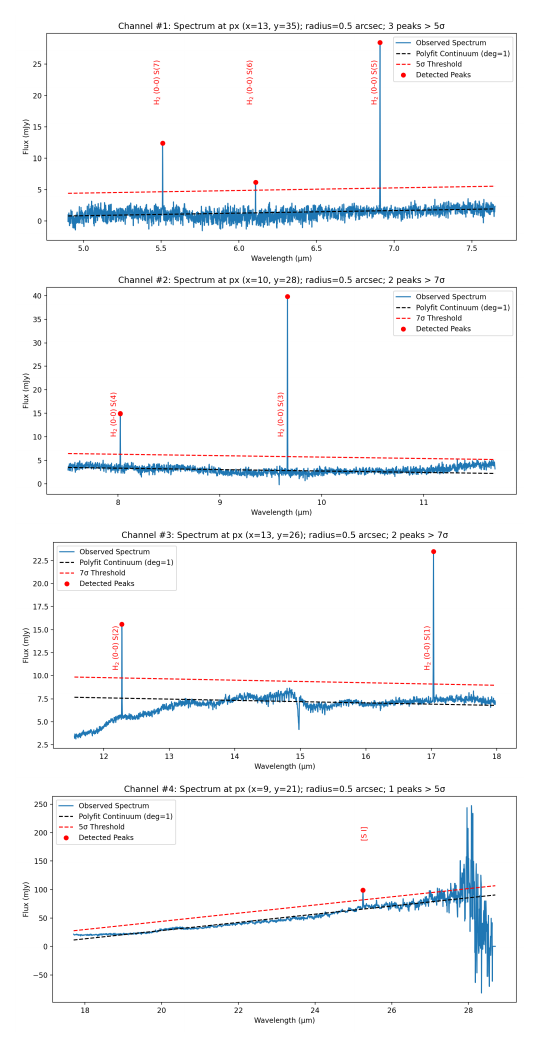}
    \caption{JWST/MIRI spectra were extracted within a 0\farcs5 radius aperture (marked with a circle in Figure~\ref{fig:spectra_ORImoment_CONTmap_ContSubtracted}) along the jet axis for all four channels. Several prominent emission features are detected across the spectral range and are marked with red dots.}
    \label{fig:peak_detection}
\end{figure}

The {\it first stage} of the analysis uses a custom Python script to extract spectra from spatially selected apertures in the JWST/MIRI IFU data cubes. The script supports both pixel-based and WCS-defined (in arcseconds) aperture extractions across the cube’s spectral axis. The extracted spectra are smoothed using a user-defined, line-free continuum window, and local maxima are identified using \texttt{scipy.signal.find\_peaks}. Peaks above a user-defined signal-to-noise threshold (typically 5-7$\sigma$) are recorded as potential emission line candidates. The output is a list of detected peak wavelengths (in $\mu$m), which can then be cross-matched with known atomic or molecular transitions.

In this paper, we focus on the prominent emission peaks; however, fainter features can also be recovered by narrowing the spectral wavelength interval, carefully selecting the continuum window, and optimizing the signal-to-noise ratio. More sophisticated filtering techniques (e.g., Savitzky–Golay filter) for the continuum fitting could enable the detection of fainter lines, but such an extension is beyond the scope of this work.
Figure~\ref{fig:peak_detection} shows the spectra from all four JWST/MIRI channels, extracted using this tool from a circular aperture of 0\farcs5 radius centered on the collimated jet (marked in Figure \ref{fig:spectra_ORImoment_CONTmap_ContSubtracted}), along the northeastern outflow axis.
Prominent peaks are successfully detected in all four channels and listed in Table \ref{tab:JWST_lines}. For Channel 4, we limited the wavelength range to $\lambda < 26~\mu$m to avoid regions contaminated by blended or noisy lines. A simple linear continuum fit (polynomial of degree 1) is applied here; however, the tool supports fitting with higher-order polynomials as needed.


The {\it second stage} takes the identified peak wavelengths from the {\it first stage} and zooms-in on each line with a small user-defined spectral window typically spanning $\pm$ 0.1$-$0.2 $\mu$m of line-free channels. The window size is adjusted based on the local line width and the proximity of neighboring features, allowing flexibility to accommodate variations in instrumental resolution and line crowding. The region immediately surrounding the line peak—excluded from the continuum fitting to avoid contamination by the line itself—is defined as a fixed-width interval centered on the peak wavelength. 
 By default, we exclude $\pm$0.014–0.02 $\mu$m around the peak, which typically corresponds to $>$ $\pm$3 spectral channels surrounding the peak channel, depending on the resolution. This width can be adjusted based on the expected line width or local spectral characteristics. The exclusion ensures that the emission feature does not bias the continuum fit. The order of the polynomial is user-defined, by default, we use a first-order (linear) polynomial fit, which we find sufficient for the relatively narrow spectral windows. To prevent overfitting, we restrict the maximum polynomial degree to 2 (quadratic). In practice, we found higher-order fits to be unnecessary given the smoothness of the continuum over these small spectral ranges. The fitted continuum is then subtracted from the original line+continuum emission to isolate the emission feature. The result is a continuum-subtracted zoom-in spectra for each detected line in the {\it first step}. First column of each row in Figure \ref{fig:spectra_ORImoment_CONTmap_ContSubtracted} are the zoom-in spectra for each identified peak.


The {\it final step} involves generating spatially resolved moment0 maps for identified emission lines. It constructs a small spectral subcube centered on each line wavelength, including both the line and nearby continuum windows similar to {\it second stage}. A pixel-wise polynomial fit of degree 1 is applied across the continuum channels to model and subtract the continuum from original emission (i.e. continuum+line), creating a continuum-subtracted cube above 3$\sigma$ of residual.  The degree of the polynomial fit is fully customizable, just as in the second stage. By default, a first-order (linear) polynomial is used, but users may adjust this up to a maximum of second order (quadratic) to prevent overfitting. From these cubes, zeroth-order moment maps (integrated intensity) are computed using the \texttt{spectral\_cube.SpectralCube} library. The output includes the original moment-0 map, a fitted continuum flux map at the peak wavelength, a continuum-subtracted moment-0 map. Both the original and continuum-subtracted moment maps have been computed using the specified channels which show emission above 3$\sigma$ of residual after continuum subtraction. These continuum-subtracted maps enable a direct spatial comparison of emission features across the field and facilitates cross-matching with ALMA maps for morphological interpretation. A similar methodology has been previously employed to generate continuum-subtracted moment maps from JWST IFU cubes \citep[e.g.,][]{2023ApJ...951L..32H,2024A&A...688A..26A,2024ApJ...966...41F,2024ApJ...962L..16N}. In our work, we have unified and streamlined this approach within the \texttt{peakmoment} tool to produce moment maps in a consistent and reproducible manner.

The second, third, and fourth panels of Figure~\ref{fig:spectra_ORImoment_CONTmap_ContSubtracted} show the original moment maps, continuum maps at the peak wavelength, and continuum-subtracted moment maps, respectively. While the outflow and jet emission is not very prominent in the original maps, the subtraction of the continuum reveals well-defined outflow structures.


\begin{figure*}[p]
    \centering
    \includegraphics[width=\textwidth]{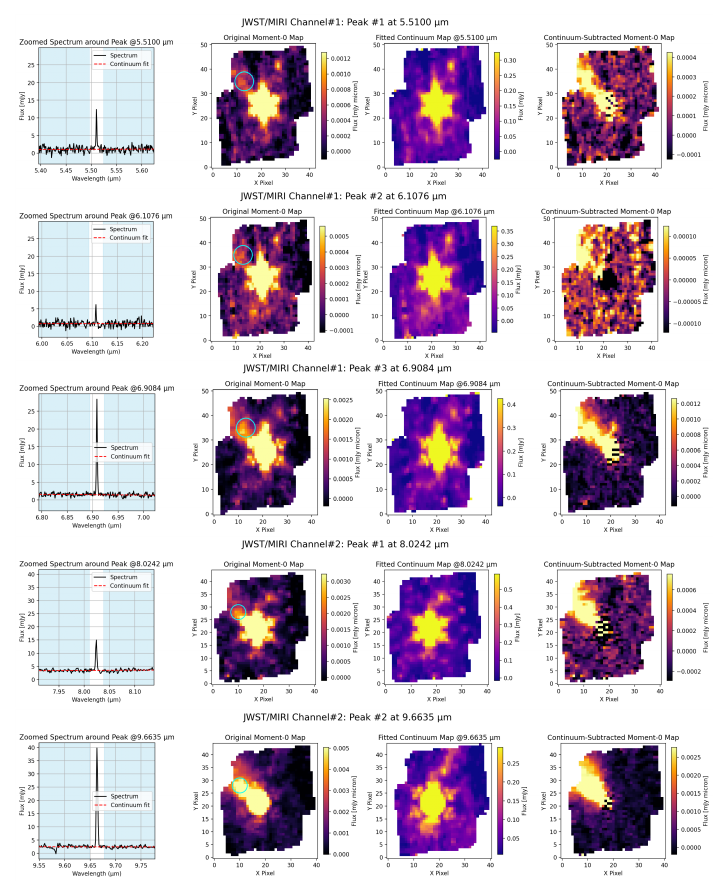}
    \caption{Continued to next page.}
\end{figure*}
\addtocounter{figure}{-1} 
\begin{figure*}[p]
    \centering
    \includegraphics[width=\textwidth]{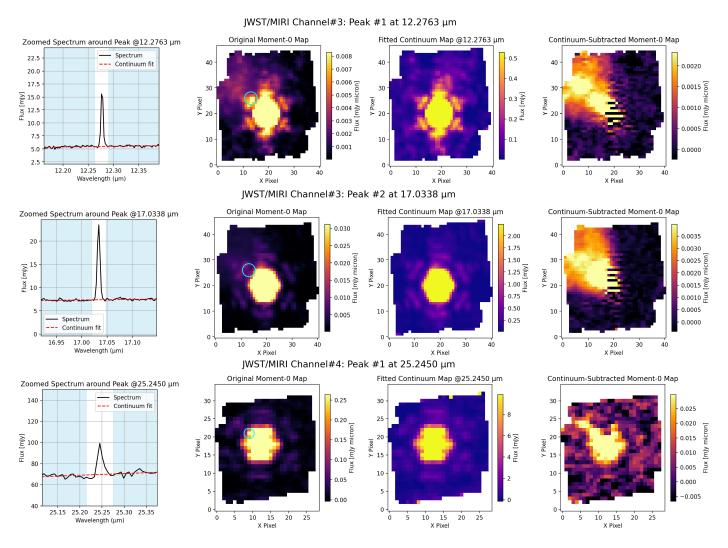}
    \caption{
    The first column shows zoomed-in spectra around each detected peak from Figure~\ref{fig:peak_detection}, with the continuum and original spectra marked. The second column displays the original moment-0 maps (before continuum subtraction) for channels with detections above 3$\sigma$ in the residual spectrum. The third column presents the continuum emission maps at the corresponding peak wavelengths. The fourth column shows the continuum-subtracted moment-0 maps for channels with detections above 3$\sigma$ in the continuum-subtracted residual spectrum. The region with a 0.5$\arcsec$ radius used for spectral extraction is indicated by a circle in the second column of each row.
}
    \label{fig:spectra_ORImoment_CONTmap_ContSubtracted}
\end{figure*}


\begin{table*}[ht]
\centering
\caption{Observed Lines from JWST/MIRI Spectrum}
\begin{tabular}{lccccc}
\hline
\textbf{Line} & \textbf{Wavelength} ($\mu$m) & \textbf{Transition} & \textbf{A$_\mathrm{ul}$ (s$^{-1}$)} & \textbf{$E_\mathrm{u}/k$ (K)} & \textbf{Notes} \\
\hline
H$_2$ 0-0S(1)     & 17.0338 & J=3 $\rightarrow$ 1   & $4.76 \times 10^{-7}$  & 1015.0  & Ortho-H$_2$ \\
H$_2$ 0-0S(2)     & 12.2788 & J=4 $\rightarrow$ 2   & $2.76 \times 10^{-6}$  & 1681.7  & Para-H$_2$ \\
H$_2$ 0-0S(3)     & 9.6635  & J=5 $\rightarrow$ 3   & $9.84 \times 10^{-6}$  & 2503.8  & Ortho-H$_2$ \\
H$_2$ 0-0S(4)     & 8.0242  & J=6 $\rightarrow$ 4   & $2.64 \times 10^{-5}$  & 3474.4  & Para-H$_2$ \\
H$_2$ 0-0S(5)     & 6.9084  & J=7 $\rightarrow$ 5   & $5.88 \times 10^{-5}$  & 4586.4  & Ortho-H$_2$ \\
H$_2$ 0-0S(6)     & 6.1076  & J=8 $\rightarrow$ 6   & $1.28 \times 10^{-4}$  & 5839.9  & Para-H$_2$ \\
H$_2$ 0-0S(7)     & 5.5100  & J=9 $\rightarrow$ 7   & $2.49 \times 10^{-4}$  & 7196.6  & Ortho-H$_2$ \\
{[S\,\textsc{i}]} & 25.2450 & $^3$P$_1$ $\rightarrow$ $^3$P$_2$ & $1.83 \times 10^{-4}$ & 396.0   & -- \\
\hline
\end{tabular}
\tablecomments{The wavelengths are the identified peaks from the JWST/MIRI IFU cubes. The line identifications and Einstein A coefficients, energy levels are adopted from \texttt{NIST Standard Atomic Spectra Database}\footnote{\url{https://www.nist.gov/pml/atomic-spectra-database}} \citep{NIST_ASD} and \texttt{Gemini important $H_2$  line list}\footnote{\url{https://www.gemini.edu/observing/resources/near-ir-resources/spectroscopy/important-h2-lines}}.}
\label{tab:JWST_lines}
\end{table*}


\subsection{Excitation Temperature Calculation from H$_2$ Lines}


To estimate the rotational temperature (\(T_{\mathrm{ex}}\)) of molecular hydrogen (H$_2$), we employed the rotational diagram (Boltzmann plot) method using the observed pure rotational transitions in the mid-infrared, as detected by JWST/MIRI. The fluxes were measured from continuum-subtracted images to properly account for extinction correction and rotational diagram analysis, as demonstrated in recent works using similar mid-infrared H$_2$ diagnostics \citep[e.g.,][]{2024ApJ...962L..16N, 2024A&A...687A..36T, 2024A&A...691A.134C}.
We extracted integrated fluxes using an aperture of  0\farcs5 radius centered on a bright ``knot B2" within the outflow/jet axis, marked with a circle in Figure \ref{fig:spectra_ORImoment_CONTmap_ContSubtracted} (see also Figure \ref{fig:ALMA_CO_SiO_cont} and \ref{fig:zoom_JWST_ALMA_comparison}). This aperture size ensures a consistent physical region is sampled across all channels despite variations at the FWHM of the PSF. It also covers at least one FWHM of the PSF at the 
 lowest-resolution channel (Channel 3) in which H$_2$ lines are detected above the 5 to 7-sigma limit.

Each rotational emission line corresponds to a transition between two rotational energy levels and is associated with a specific upper level energy $E_u$, degeneracy $g_u$, and Einstein $A_{ul}$ coefficient (Table \ref{tab:JWST_lines}). Assuming local thermodynamic equilibrium (LTE), the population of the upper levels follows the Boltzmann distribution:

\begin{equation}
\ln\left(\frac{N_u}{g_u}\right) = \ln\left(\frac{N_{\mathrm{tot}}}{Z(T)}\right) - \frac{E_u}{kT_{\mathrm{ex}}}
\end{equation}

where: $N_u$ is the column density of the upper level, $g_u$ is the statistical weight of the upper level, $E_u$ is the excitation energy of the upper level (in K), $k$ is the Boltzmann constant, $T_{\mathrm{ex}}$ is the excitation temperature, $Z(T)$ is the partition function.
\begin{figure}
    \centering
    \includegraphics[width=1\linewidth]{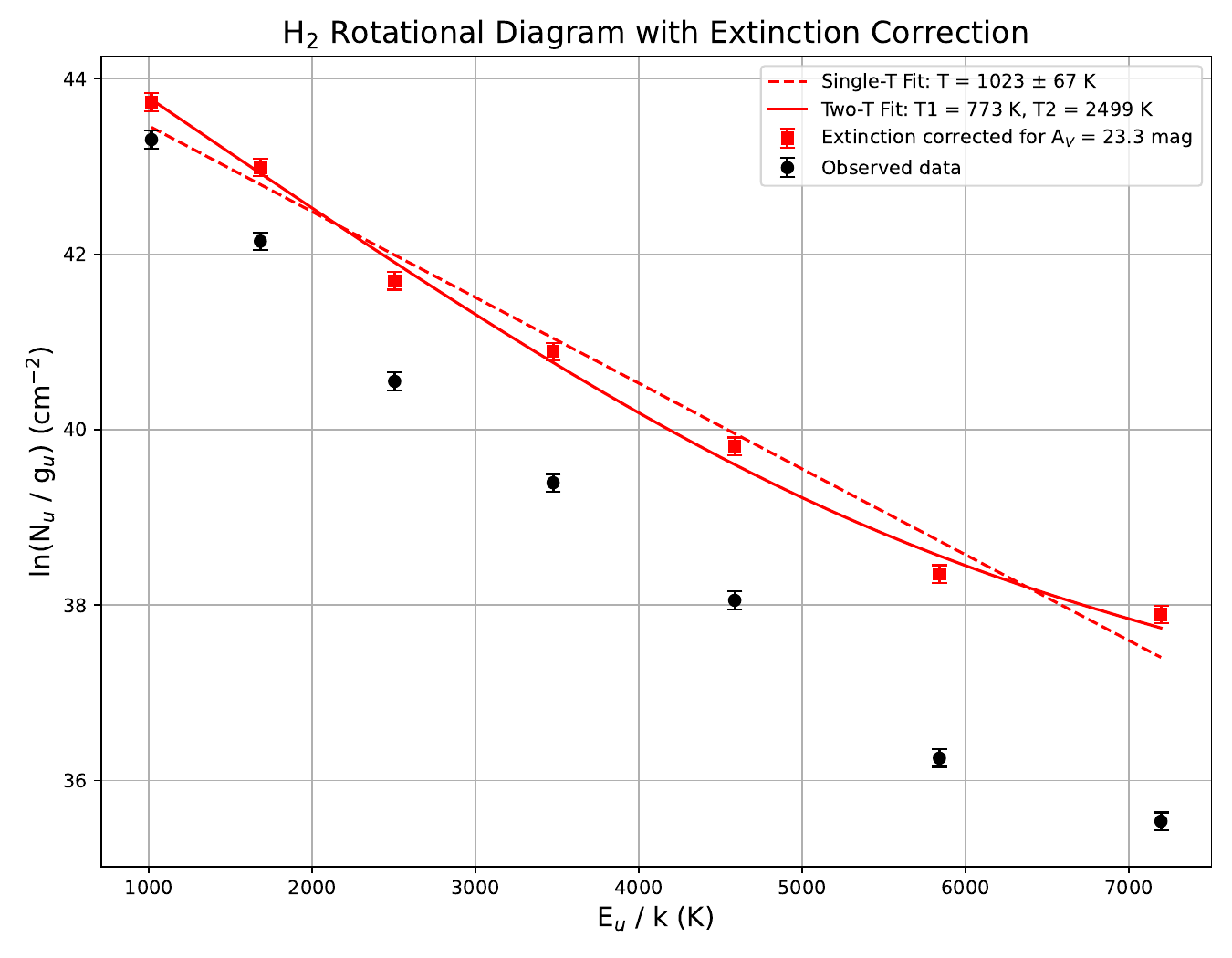}
    \caption{Rotational diagram of H$_2$ pure rotational transitions observed with JWST/MIRI, based on integrated emission line fluxes. The logarithm of the column density per statistical weight ($\ln(N_u/g_u)$) is plotted against the upper-level energy ($E_u/k$) for selected transitions. Black points represent values derived from the observed fluxes, while red points are corrected for extinction. The red dotted line indicates a single-temperature fit, and the curved red line represents a two-temperature component fit.
}
    \label{fig:rotation_diagram}
\end{figure}
Assuming the $H_2$ lines are optically thin, the upper level column density is related to the observed line flux $F_\lambda$ by:

\begin{equation}
N_u = \frac{4\pi F_\lambda}{A_{ul} h \nu \Omega}
\end{equation}

Here, $\nu$ is the frequency corresponding to the wavelength $\lambda$, $h$ is Planck’s constant, and $\Omega$ is the solid angle over which the emission is integrated. 
To derive $T_{\mathrm{ex}}$, we plot $\ln(N_u / g_u)$ against $E_u$ in Figure \ref{fig:rotation_diagram}. A linear fit to this plot gives a slope of $-1/T_{\mathrm{ex}}$. 




\subsubsection{Extinction Correction in H$_2$ Rotational Diagram Analysis}\label{sec:extinction_corrected_rotational_diagram}

The observed mid-infrared fluxes of molecular hydrogen (H$_2$) rotational lines are affected by interstellar dust extinction, particularly in the 5--20\,$\mu$m range. To recover the intrinsic line intensities and accurately derive the excitation conditions of the gas, it is essential to correct the measured fluxes for extinction.

To estimate the line-of-sight visual extinction ($A_V$), we compare the observed flux (i.e. non-continuum subtracted) ratios of H$_2$ 0-0 pure rotational transitions with their expected intrinsic values under LTE conditions at different excitation temperatures. We have estimated an mean visual extinction A$_V$ = 23.3$\pm$2.5 mag from the observed integrated line flux ratios of ortho-to-para. This visual extinction along the jet axis near the continuum peak is consistent with the high-extinction environments around embedded protostars.

In our analysis, we adopted this estimated visual extinction value, and used the \citet{2001ApJ...548..296W} extinction curve for the mid-infrared regime to correct extinction to the observed fluxes.  We applied a rotational diagram analysis, corrected for extinction, to derive the excitation conditions of H$_2$. A single-temperature fit yields a rotational temperature of 1023 $\pm$ 67\,K, consistent with a warm molecular component. To account for the observed curvature in the excitation diagram, we also performed a two-temperature fit, resulting in T$_\mathrm{1}$=773$\pm$44\,K and T$_\mathrm{2}$=2499$\pm$721\,K. To evaluate the influence of the dust extinction model, we repeated both the single- and two-temperature fits using the icy dust model KP5 \citep[][]{2024RNAAS...8...68P} and the dense-cloud model of \citet[][]{2009ApJ...693L..81M}. The derived temperatures vary within the respective uncertainties; however, the overall two-component structure and temperature trend remain consistent.

The longer-wavelength transitions (e.g., S(1)$–$S(4)) primarily trace cooler gas, while the shorter-wavelength lines (e.g., S(5)$–$S(7)) probe hotter regions. Such two temperature environment was also observed in the jets/outflows of several other protostellar objects \citep[e.g.,][]{2024ApJ...962L..16N,2025ApJ...982..149O}. This trend is consistent with longer-wavelength lines becoming optically thick in dense, shock-heated regions, limiting their sensitivity to warm gas, whereas shorter-wavelength lines remain optically thin and trace deeper, hotter gas \citep[e.g.,][]{2006ApJ...649..816N, 1992ApJ...399..563B}. The results suggest a stratified structure in the emitting region, likely due to a combination of high-density molecular shocks and UV-irradiated outflow cavity walls. The presence of both warm and hot components indicates emission from a mixture of thermal environments, including shock-heated gas and UV-excited photodissociation regions (PDRs). 
Utilizing two independently estimated fits, we derive column densities of 
\( N_1 = (3.34 \pm 0.43 \times 10^{20}~\mathrm{cm}^{-2} \) for temperature component \( T_1 \), and 
\( N_2 = (1.11 \pm 1.03) \times 10^{19}~\mathrm{cm}^{-2} \) for \( T_2 \). 
The combined mean molecular hydrogen column density is estimated to be 
\( N(\mathrm{H}_2) = (1.72 \pm 0.22) \times 10^{20}~\mathrm{cm}^{-2} \).

\subsection{Combining ALMA Maps with JWST Observations}

\begin{figure*}
    \centering
    \includegraphics[width=1\linewidth]{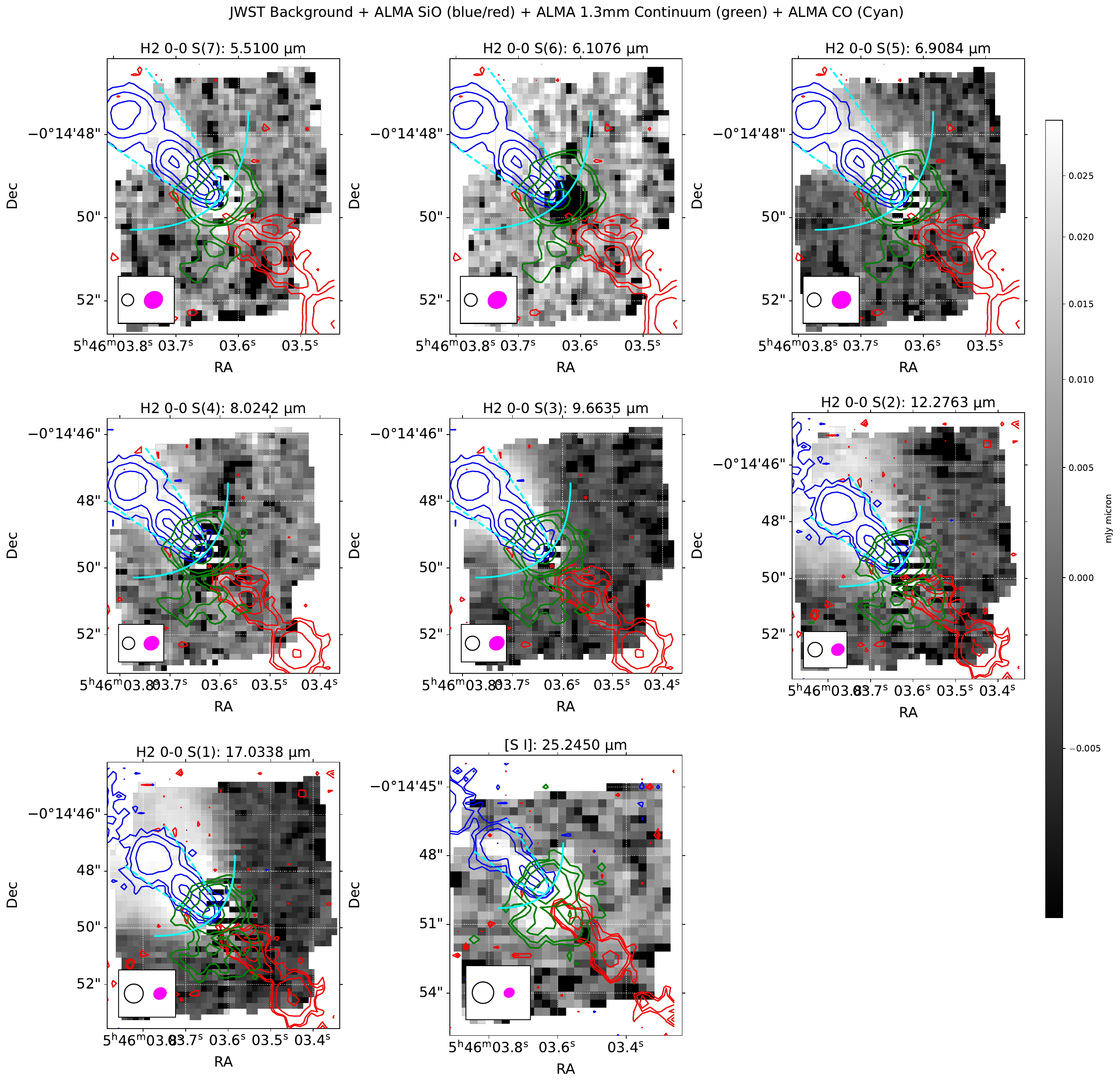}
    \caption{
    Comparison of infrared JWST/MIRI and submillimeter ALMA Band~6 observations.  Blue and red contours represent the ALMA SiO (5--4) integrated emission at the [90, 94, 97, 99.5] percentiles, with a sensitivity of 7~K~km~s$^{-1}$, overlaid on the JWST/MIRI H$_2$ maps.
     The dotted sky-blue parabola indicates the approximate high-velocity CO outflow morphology traced by $>$3$\sigma$ sensitivity emission. The solid parabola with larger curvature outlines the low-velocity CO outflow cavity, also detected above the 3$\sigma$ sensitivity level. Green contours represent the [90, 94, 97, 99.5] percentiles of the ALMA Band~6 continuum map, which has a sensitivity of 0.8~mJy~beam$^{-1}$. The JWST/MIRI PSF FWHM (wavelength-dependent) is shown as a black ellipse in the lower left corner, and the ALMA Band~6 synthesized beam is indicated in magenta.
}
    \label{fig:JWST_ALMA_comparison}
\end{figure*}

Figure~\ref{fig:ALMA_CO_SiO_cont} shows the ALMA observations of CO and SiO emission. In Figure~\ref{fig:JWST_ALMA_comparison}, the background image presents continuum-subtracted moment maps from JWST. The SiO emission from ALMA is overlaid as contours on top of the JWST images. Since the JWST resolution varies with wavelength, the PSF FWHM is estimated using the relation \(\theta_{\mathrm{FWHM}} = 0.033\,(\lambda/\mu\mathrm{m}) + 0.106\arcsec\) from \citet{2023AJ....166...45L}, which yields approximately \(0\farcs29\) at \(\lambda = 5.51\,\mu\mathrm{m}\) and \(0\farcs67\) at \(\lambda = 17.0338\,\mu\mathrm{m}\). These PSF sizes are illustrated as black circles in Figure~\ref{fig:JWST_ALMA_comparison}. The average ALMA synthesized beam, approximately \(0\farcs46 \times 0\farcs40\), is shown in magenta. To enable spatial comparison between JWST and ALMA, we resampled the ALMA maps onto the JWST pixel grid using the \texttt{reproject.reproject\_interp} function, which applies bilinear interpolation to estimate values at the new pixel locations. This process aligns the coordinate systems without altering the intrinsic beam resolution. We did not convolve the maps to match beam sizes, as doing so would require degrading all datasets to the lowest resolution—here, that of JWST/MIRI Channel~3. Such convolution would result in a loss of morphological detail, particularly the jet ``knots," which are narrower than the beam sizes and are critical for tracing fine-scale jet structures.
Due to the differing spatial resolutions, any morphological comparison between ALMA and JWST should be interpreted with caution. 
 For example, the JWST/MIRI maps with lower angular resolution than the ALMA maps may cause compact features to appear more extended or blended. These resolution differences can lead to apparent discrepancies in morphology or alignment that do not reflect true physical structures. 


 As demonstrated in Appendix \ref{sec:Appendix_ALMA_cosiocont} and Figure \ref{fig:Appendix_ALMA_cosiocont}, 
the CO maps reveal two distinct components: one at higher velocities ($|V_{observed}-V_{systemic}| > 64\ km\,s^{-1}$) and another at lower velocities ($0 < |V_{observed}-V_{systemic}| < 64\ km\ s^{-1}$) \citep[see also,][]{2022ApJ...925...11D}. The high-velocity CO component closely resembles the morphology of the SiO emission, observed only in higher velocities ($|V_{observed}-V_{systemic}| > 64\ km\,s^{-1}$). In Figure~\ref{fig:JWST_ALMA_comparison}, the high-velocity CO boundary is outlined with a dashed parabolic curve, concentrated primarily along the jet axis and coinciding with the SiO emission region. In contrast, the low-velocity CO emission displays a broader, wide-angle parabolic structure, indicative of a wider outflow cavity. Note that the SiO contour levels vary slightly across JWST/MIRI maps at different wavelengths due to changes in pixel scale and interpolation effects during reprojection of ALMA data onto the JWST pixel grid.



\section{Discussion}

\subsection{Comparison of infrared-JWST  with submillimeter-ALMA emission}
The advent of ALMA and JWST MIRI marks a transformative era in protostellar jet and outflow studies. By combining ALMA’s submillimeter sensitivity to cold and warm molecular gas (e.g., CO, SiO) with MIRI’s access to warm and hot gas via mid-infrared atomic and molecular lines, we gain a comprehensive, multi-phase view of jet structure, dynamics, and origin. ALMA maps dense, shocked, and entrained material from a few to thousands of AU, resolving molecular jets, outflows, and disk winds with precision. In contrast, JWST MIRI probes the inner hot molecular, ionized and atomic regions of jets—within just a few AU of the protostar—unreachable by ALMA due to high temperatures and ionization. Together, these observatories reveal the spatial coexistence of ionized, atomic, and molecular components, highlighting stratified launching zones. 

ALMA submillimeter emission in $^{12}$CO (2-1) typically traces both the wide-angle, low-velocity wind components and the high-velocity collimated  jet components in protostellar outflows.   In contrast, SiO emission is more selectively associated with the high-velocity, collimated jet component. 
ALMA SiO (5-4), however, is often enhanced in strong shocks where dust grains are sputtered, releasing silicon into the gas phase \citep[][]{1991ApJ...373..254G,1997A&A...321..293S,2017NatAs...1E.152L,2024AJ....167...72D}. As a result, SiO highlights localized, shock-excited regions within the jet.


JWST observations show that the shorter-wavelength H$_2$ lines (e.g., S(7), S(6), S(5), and S(4)) are more sensitive to emission concentrated along the jet axis, typically tracing the hotter and more collimated jet components.
As the wavelength increases, the emission becomes progressively more extended and wide-angled, reflecting contributions from cooler, entrained molecular gas. At longer wavelengths-S(3), S(2), and S(1)-the H$_2$  emission traces both the bright axial jet and the broad, wide-angle outflow cavity, revealing the full structure of the protostellar outflow.

 Figures  \ref{fig:JWST_ALMA_comparison} and \ref{fig:zoom_JWST_ALMA_comparison} demonstrate that the shorter-wavelength H2 lines in the JWST/MIRI bands trace similar compact, collimated jet structures to the high-velocity SiO (5-4) emission observed with ALMA. In contrast, the longer-wavelength H$_2$ lines (S(3), S(2), and S(1)) trace broader, more diffuse outflow components, while still capturing inner jet features that resemble the morphology seen in low-velocity $^{12}$CO (2-1) emission. SiO and CO have upper-level excitation temperatures of approximately $\sim31.26$~K and $\sim16.60$~K, respectively, indicating that they trace the cooler components of the jet environment, although both species can survive at temperatures of several hundred kelvin without dissociation. In contrast, the infrared H$_2$ transitions detected with \textit{JWST}/MIRI probe significantly hotter gas, with excitation temperatures ranging from $\sim1000$~K to $7000$~K (see Table~\ref{tab:JWST_lines}), and potentially even higher in NIRSpec detections \citep[e.g.,][]{2025A&A...695A.145V}. These characteristics could be the indication of the presence of stratified layers of gas within the jet, detected based on their varying excitation temperatures and molecular abundances. To fully characterize this structure, high-sensitivity, multiwavelength observations, combined with theoretical modeling of radial temperature gradients, will be essential.


The emissions observed with JWST and ALMA are summarized in the schematic diagram shown in Figure~\ref{fig:schematic}. The outflow cavity is traced by the $H_2$ 0--0 S(1) and S(2) lines, along with CO(2--1) emission. The $H_2$ 0--0 S(3) line highlights slightly denser components along the jet axis. The $H_2$ 0--0 S(4) and S(5) transitions trace dense gas layers, consistent with the high-velocity zones also seen in CO(2--1). The $H_2$ 0--0 S(6), S(7), and the ionized [S~\textsc{i}] line are particularly sensitive to the narrow, high-temperature jet components.

\subsubsection{Shock Structure}

\begin{figure*}
    \centering
    \includegraphics[width=1\linewidth]{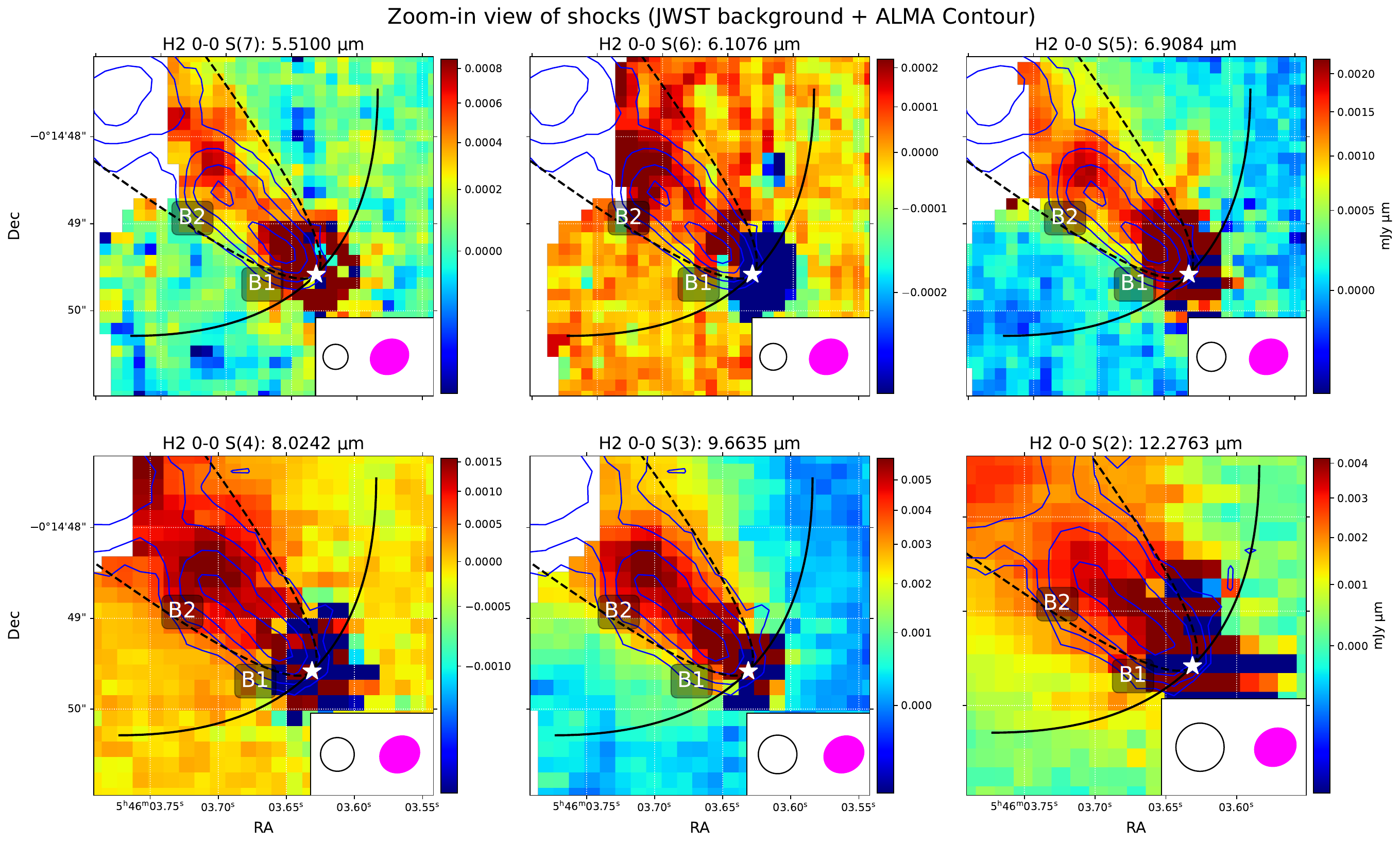}
    \caption{
    Zoom-in view of the JWST and ALMA observations highlighting the shock structures in greater detail. The blue contours, as well as the dotted and solid parabolas, are the same as in Figure~\ref{fig:JWST_ALMA_comparison}, representing the SiO (5--4) emission and the high- and low-velocity CO outflow structures, respectively. Two shock knots, labeled B1 and B2, are marked. The position of the protostar is also indicated. The JWST/MIRI PSF FWHM and the ALMA Band~6 synthesized beam are shown in the lower right corner.
}
    \label{fig:zoom_JWST_ALMA_comparison}
\end{figure*}

Figure~\ref{fig:zoom_JWST_ALMA_comparison} presents a zoomed-in view of the knots. Using ALMA SiO and CO observations, \citet[][]{2022ApJ...925...11D} reported a jet velocity of 125~km\,s$^{-1}$ for G205S3, assuming an inclination angle of 40$\degr$. Consistent with this, \citet[][]{2025A&A...695A.145V} derived a terminal jet velocity of 120-125~km\,s$^{-1}$ based on JWST infrared observations. Considering these velocities, the inclination angle, and the distance to Orion, the dynamical age of knot B1 is estimated to be $\sim$10 years at a projected separation of $\sim$ 210~AU, while knot B2 is $\sim$30 years old at $\sim$ 630~AU. These two knots thus represent newly ejected material still within the protostellar envelope, undergoing shock processing.

A noticeable northward shift (sideway to the jet) is observed in the shocked knots traced by the $H_2$ emission at shorter wavelengths compared to the SiO emission.  
 The positional offset between the JWST/MIRI H$_2$ emission at shorter wavelengths and the ALMA SiO Gaussian centroid of knot B2 ranges from $\sim$ 0\farcs45 at 5.51 $\mu$m to 0\farcs26 at 9.665 $\mu$m. At longer JWST/MIRI H$_2$ wavelengths, the knot emission appears increasingly diffuse, making it challenging to isolate and fit a distinct component. 
Several factors could contribute to this offset. One possibility is the difference in beam sizes between JWST and ALMA, which can artificially shift the apparent emission peaks, complicating the interpretation of the intrinsic morphology of the knots. In particular, JWST's PSF is more circular, while ALMA's synthesized beam is elliptical, potentially introducing asymmetries in spatial distribution. However, the shift in the peak position of the knots persists even when the beam sizes are comparable, such as in the case of the $H_2$ 0--0 S(4) transition at 8.0242~$\mu$m.  Another contributing factor could be a slight mismatch in the World Coordinate System (WCS) solutions of the two instruments. The alignment of ALMA and JWST images depends on the accuracy of their respective WCS calibrations, and even small discrepancies can introduce positional offsets that affect the interpretation of multi-wavelength structures.

Another plausible explanation is the system’s viewing geometry. Given the inclination angle of approximately $40\degr \pm 8\degr$, if the observer's viewpoint is from the northeast, they would predominantly see the upper, cooler layers traced by CO(2--1) and SiO(5--4), whereas the hotter, inner jet traced by $H_2$ emission would appear spatially offset due to projection effects. This viewing angle also supports the observed asymmetry in the outflow: a bright, blueshifted lobe is detected in $H_2$ emission, while the redshifted counterpart is likely obscured by envelope extinction.

\begin{figure*}
    \centering
    \includegraphics[width=0.7\linewidth]{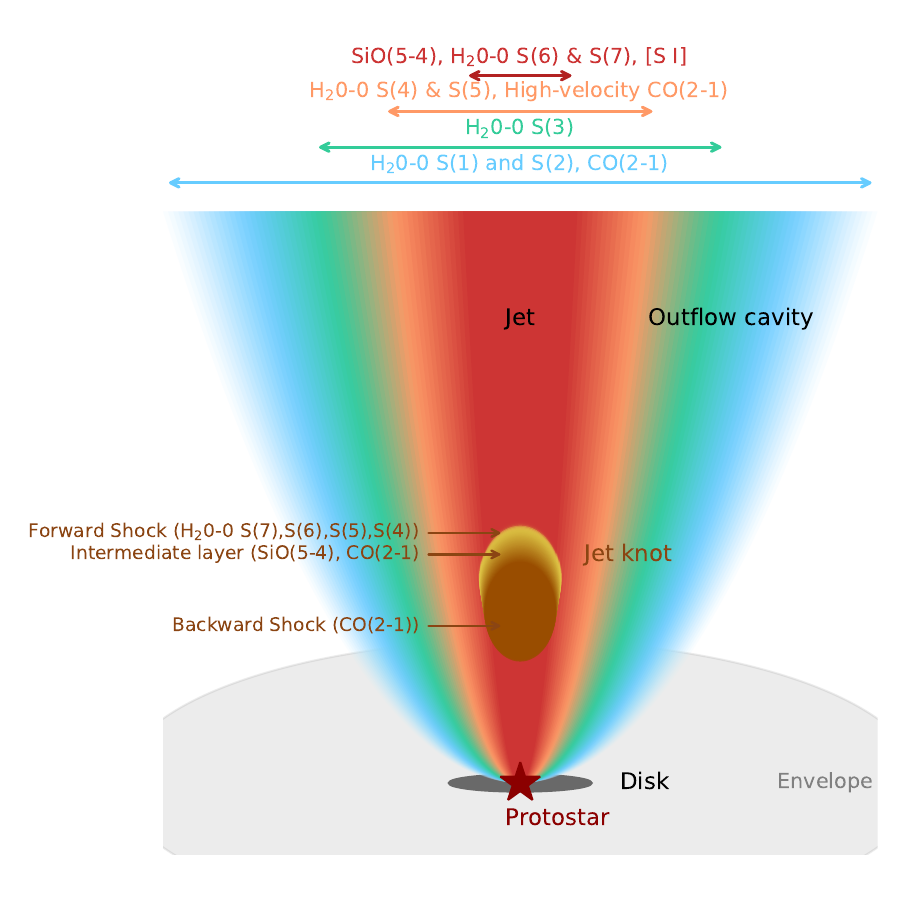}
    \caption{Schematic diagram of the jet/outflow system. The upper section indicates the spatial ranges of the outflow and jet detected with JWST and ALMA. A representative knot is illustrated along the jet axis. Distinct emission layers traced by different spectral lines observed with ALMA and JWST are labeled, highlighting the stratified structure of the shock regions.}
    \label{fig:schematic}
\end{figure*}

Another important aspect is that the knot peaks (B2 in Figure \ref{fig:zoom_JWST_ALMA_comparison}) in H$_2$ emission lines are systematically shifted toward the North-East along the jet axis than SiO emission peak. This may indicate that the $H_2$ emission is tracing the high-temperature forward layer of the shock front, whereas SiO and CO trace intermediate-temperature regions. The cooler, backward layer is primarily traced by CO. Based on our findings and comparisons with previous shock models \citep[][]{2001ApJ...557..429L,2009A&A...495..169S,2017A&A...597A.119T,2021ApJ...909...11J,2024AJ....167...72D}, we present a schematic diagram in Figure \ref{fig:schematic} illustrating the shock structure as it precesses along the jet axis.
In this framework, the forward collision zone corresponds to the hottest shocked layer, best traced by short-wavelength $H_2$ transitions (S(7) to S(4)), which is consistent with the high excitation temperatures derived from the rotational diagrams. The SiO(5--4) emission likely traces a slightly cooler post-shock region immediately behind the hottest layer. The backward shock is the coolest among the three, traced by CO(2--1). 
The innermost region within the jet in the backward layer, however, may still be at very high temperatures and is partially traced by $H_2$, although it appears fainter than the forward layer. In contrast, the CO(2--1) emission primarily represents the cooler, outer surrounding the hotter $H_2$ jet.

The ALMA observations were conducted on 1 November 2018, followed by JWST observations on 15 March 2023, with a time gap of approximately 4.37 years. Assuming a typical proper motion corresponding to a mean jet velocity of $125~\mathrm{km\,s^{-1}}$, the expected displacement over this period is $\sim115$~AU (equivalent to $\sim0.22\arcsec$ at Orion distance 400 pc and inclination angle 40$\degr$). This expected shift is approximately half of the observed positional offset between the Gaussian peak at the ALMA SiO  and the JWST/MIRI H$_2$ emission at 5.5\,$\mu$m for knot B2, suggesting that proper motion alone may not fully account for the observed displacement. However, the expected shift remains close to the angular resolution limits of the data, that is the FWHM of the JWST PSF and the ALMA beam. Therefore, the current spatial resolution and temporal baseline are insufficient to robustly confirm or rule out proper motion based solely on peak position offsets. Future multi-epoch observations with comparable or improved spatial resolution at the same wavelengths are required to reliably constrain knot motions.
 
\subsubsection{Monopolar in infrared but bipolar in submillimeter}
Interestingly, the blueshifted jet is prominently detected only in the JWST/MIRI data, while its counterpart redshifted is notably absent, despite both red- and blueshifted components being identified in ALMA SiO and CO observations. However, asymmetry is also seen in ALMA SiO emission;   the redshifted emission is fainter than that of the blueshifted jet, as evident from SiO maps in Figures \ref{fig:ALMA_CO_SiO_cont} and \ref{fig:Appendix_ALMA_cosiocont} (panel 3) \citep[see also,][]{2022ApJ...925...11D}. Thus, the apparent absence of the redshifted jet in the infrared does not confirm its true absence. A comprehensive multiwavelength investigation is essential to uncover the full morphology and structure of the protostellar jet and outflow system.


The observed asymmetry in the jet emission may arise from multiple contributing factors, with geometry playing a particularly important role. In protostellar systems, infrared emission detected by JWST can be significantly affected by extinction from the surrounding envelope, especially along the redshifted side of the flow \citep[e.g.,][]{2023ApJ...951L..32H,2024A&A...692A.143B,2024ApJ...966...41F,2025A&A...699A.361V}. 
Given the inclination angle of approximately $40\degr$, if the observer is positioned toward the blueshifted lobe, the redshifted emission near the base of the jet must traverse a longer path through the dense, dusty envelope. This results in greater extinction and a fainter observed redshifted lobe in the infrared. The spatial orientation of the jet relative to the line-of-sight can also introduce projection effects, further contributing to the apparent asymmetry. Regions of the jet that are still embedded within the envelope may remain undetected due to high optical depth.


Doppler boosting can cause brightness asymmetries in relativistic jets—such as those observed in AGNs \citep[e.g., M\ 87,][]{2007ApJ...668L..27K,2007ApJ...658..232C} but this effect is unlikely to play a significant role in protostellar jets due to their much lower velocities. Although protostellar jets can reach velocities of a few hundred km\,s$^{-1}$, these speeds are insufficient to produce noticeable Doppler boosting.  Interestingly, observations show that nearly 50\% of protostars with SiO jets appear monopolar in millimeter wavelengths \citep[][references therein]{2024AJ....167...72D}.  Extinction alone is unlikely to fully explain this high incidence of monopolar jets observed at submillimeter wavelengths, where dust opacity is relatively low and extinction effects are minimal. This suggests the possibility of an intrinsic launching asymmetry, perhaps arising from the underlying magnetic field structure or star–disk interactions, that may favor the formation of unipolar jets in certain systems.

In the models of jet-launching, when a stellar magnetosphere is included, it introduces oppositely directed poloidal fields around the rotation axis. This leads to a more complex quadrupolar toroidal field structure, allowing reconnection and jet-driving mechanisms to operate in both hemispheres—supporting symmetric, bipolar jets \citep[e.g.,][]{1994ApJ...429..781S,2000prpl.conf..789S,2003ApJ...599..363A,2007prpl.conf..277P}.  

In contrast, in jet-launching models where only the disk contributes to the magnetic field (i.e., no stellar magnetosphere), the structure of the magnetic field becomes inherently asymmetric. The toroidal magnetic field in such systems typically forms a unipolar configuration within each hemisphere, meaning there is no reversal of field direction across latitudes. As a result, magnetic reconnection and amplification processes, such as avalanche accretion streams that are necessary for strong jet launching, can only develop effectively in one hemisphere—specifically, where the toroidal field reverses \citep[][]{2025ApJ...988..107T,2025arXiv250611333T}.
This asymmetry leads to conditions favorable for launching a jet on only one side of the system, while the opposite lobe remains weak or entirely suppressed. Hence, the system preferentially produces a unipolar jet rather than a symmetric bipolar outflow.

\subsection{Implication of Atomic Sulphur emission} 
The [S\,\textsc{i}] 25.24\,$\mu$m emission near the source appears to originate from ionized material in the surrounding region, including both the immediate environment of the source and the jet itself (Figure~\ref{fig:JWST_ALMA_comparison}, last panel). In addition to that, due to the limited spatial resolution, knots B1 and B2 are not clearly distinguishable. The [S\,\textsc{i}] emission is extended along the jet axis. Although it has a larger beam size than H$_2$ S(7) at 5.5 $\mu$m, the overall morphology appears similar. This emission also resembles the ALMA SiO jet structure, suggesting that it traces ionized components along the jet axis. However, the lower spatial resolution at longer wavelengths for [S\,\textsc{i}] limits a direct comparison of the knot structures.

The [S\,\textsc{i}] fine-structure line is a powerful tracer of neutral atomic sulfur in warm, dense photodissociation regions (PDRs), occurring near the interface between ionized and molecular gas. In such environments, far-ultraviolet photons dissociate molecules and ionize atoms, yet sulfur can remain largely neutral within layers where hydrogen is molecular while carbon is ionized \citep{2024A&A...687A..87F}. 

The [S\,\textsc{i}] line also serves as a key diagnostic of shock-excited gas in protostellar collimated jet (Figure \ref{fig:JWST_ALMA_comparison}, last panel). Shock processes can dissociate or sputter sulfur-bearing molecules (e.g., SO, CS, H$_2$S), releasing atomic sulfur into the gas phase \citep{2009ApJ...706..170N}, and the resulting [S\,\textsc{i}] emission probes post-shock chemistry, energetics, and elemental depletion in the jets.

\subsection{Jet Physical condition}

As summarized in Figure \ref{fig:schematic} based on observations, the outflow cavity to inner jet could have different stratified layer. Zooming into the jet axis and knot structure, two component of temperatures has been estimated in this study, while JWST/MIRI shorter  wavelengths sensitive to hotter gas than longer wavelengths.  We have estimated two component fit to the rotational diagram, suggesting two layer of temperature, one components with 
 $\sim 773 \pm 44~K$ another with  $\sim 2499\pm720~K$.
\citet[][]{2025A&A...695A.145V} also reported two different gas components for G205S3 object cooler gas temperatures of 500–600\ K from mid-infrared JWST/MIRI and a range of temperature 2400–3800\ K from JWST/NIRSpec. Similar two component jet temperatures have been reported in other sources such as IRAS 16253-2429 \citep[][]{2024ApJ...962L..16N}, IRAS 15398-335 \citep[][]{2025ApJ...982..149O}. 




To estimate the mass flux through the jet, we model the inner jet as a cylindrical flow of gas primarily composed of molecular hydrogen, with a mean column density \( N(\mathrm{H}_2) \) (see Section~\ref{sec:extinction_corrected_rotational_diagram}) and a constant jet velocity \( V_j = 125~\mathrm{km\,s^{-1}} \). The mass-loss rate through the jet can be approximated by the following expression:

\begin{equation}
    \dot{M_J} = \mu_{H_2} m_{\mathrm{H}} N(\mathrm{H}_2) V_j W_j
\end{equation}

where \( \mu_{H_2} = 2.8 \) is the mean molecular weight per hydrogen molecule, \( m_{\mathrm{H}} \) is the mass of a hydrogen atom, and \( W_j \) is the width of the jet perpendicular to the flow direction. Since the jet is unresolved at the current spatial resolution -its width is expected to be much smaller \citep[][]{2017NatAs...1E.152L}- we adopt the highest FWHM of the JWST/MIRI PSF at 5.5~\micron, approximately \( 0\farcs29 \), as an estimate for \( W_j \).

We estimate a jet mass-loss rate of \(\dot{M}_{\rm J, blue, JWST} \approx 0.27 \times 10^{-6} \pm 0.2 \times 10^{-6}~M_\odot~\mathrm{yr}^{-1}\), consistent with previous values derived from ALMA CO observations of the blueshifted lobe alone, i.e., \(\dot{M}_{\rm J, blue, ALMA} \approx 0.29 \pm 0.12 \times 10^{-6}~M_\odot~\mathrm{yr}^{-1}\) \citep{2022ApJ...925...11D}. Although the redshifted lobe is not detected in the present JWST/MIRI data, it is clearly visible in submillimeter observations. Therefore, the total mass-loss rate can be approximated as roughly twice that of the blueshifted component i.e. \(\dot{M}_{\rm J, total, JWST} \approx 0.54 \pm 0.4 \times 10^{-6}~M_\odot~\mathrm{yr}^{-1}\) which is in agreement with that of estimated from ALMA observation \(\dot{M}_{\rm J, total,ALMA} \approx 0.59 \pm 0.24 \times 10^{-6}~M_\odot~\mathrm{yr}^{-1}\).  This level of mass loss suggests that the protostar is undergoing a moderately high accretion phase, similar to those seen in well-studied Class~0 sources such as HH~212 and HH~211 \citep[][]{2020A&ARv..28....1L}. Such high mass-loss rates enhance the density within the jet, making it particularly favorable for detection in high-density tracers such as millimeter SiO~(5--4), which requires a critical density of \(\sim \mathrm{1.7 \times 10^{6} ~cm^{-3}}\) at a jet temperature of 150~K \citep{2021A&A...648A..45P,2024AJ....167...72D}.


\subsection{Source Evolutionary Status and its possible implication on planet formation}

Following the classification criteria based on bolometric temperature and spectral index \citep{1995ApJ...445..377C,2009ApJ...692..973E}, the SED of G205S3 suggests that it is a Class~I protostellar system. It is characterized by a bolometric temperature of \(T_{\mathrm{bol}} = 180 \pm 33~\mathrm{K}\) and a spectral index of \(\alpha \sim 0.417\) \citep[see HOPS\,315;][]{2016ApJS..224....5F,2020ApJS..251...20D}. Despite this classification, G205S3 exhibits an energetic, collimated jet traced by molecular SiO in submillimeter emission, along with a moderately high mass-loss rate, estimated both from infrared and submillimeter, these features more typical of well-known Class~0 protostars. These seemingly contradictory characteristics raise questions about the evolutionary status of the source.  Therefore, the source G205S3 may be in the earliest phase of Class~I evolution \citep[see also][]{2025Natur.643..649M}, still experiencing elevated accretion activity. Alternatively, it could represent a young Class~0 source that appears more evolved due to observational geometry \citep[e.g.,][]{2023ApJ...944...49F}. Based on CO outflow morphology, \citet[][]{2022ApJ...925...11D} estimated an inclination angle of \(\sim 40^\circ \pm 8^\circ\), which would allow a more direct view of the central protostar compared to a typical edge-on system. From the perspective of the observer, located in the Solar System, the system is likely viewed from its north-eastern direction, looking down the blueshifted lobe to the protostellar system. This geometry may also explain why the continuum emission (green contours) appears slightly shifted toward the blueshifted side in Figures~\ref{fig:ALMA_CO_SiO_cont} and \ref{fig:JWST_ALMA_comparison}. Such a relatively unobscured, pole-on-like viewing angle can enhance the observed infrared brightness, potentially leading to an overestimation of the bolometric temperature and, consequently, a misclassification of the evolutionary stage— since the SED-based classification is highly sensitive to line-of-sight extinction and inclination \citep[][]{1994ApJ...434..330C,2016ApJS..224....5F}.


If the source is indeed an evolved Class~I  as classified from SED or even in a Class~0 system, these  evolutionary stages represent the critical windows for planet formation, particularly in the outer regions of the protostellar disk where icy planetesimals, asteroids, and comets are expected to form \citep[e.g.,][]{2014prpl.conf..339T,2018ApJ...869L..46D}. Recent studies of the G205S3 protostar by \citet[][]{2025Natur.643..649M} suggest the source could be at an early Class~I evolutionary stage, and have revealed new insights into this early planet-forming environment using JWST and ALMA observations.  They detected warm SiO gas and crystalline silicate minerals within $\sim$2.2 AU of the source, spatially separated from the SiO jet. These silicates are interpreted as refractory solids recondensed from vaporized interstellar grains, marking the onset of planet formation analogous to the earliest solid formation events in the Solar System. The inclination of the system allowed JWST to detect hot crystalline silicate emission arising from the upper layers of the inner disk, rather than from an outflow or scattered light, thereby enabling localization of the silicate origin to the disk itself.

Feedback due to the the elevated mass ejection through the jet in this  G205S3 protostellar system  can inject significant momentum and heat into the surrounding environment, enhancing the hot gas component in the outer disk's surface layers \citep[][]{2021ApJ...920L..35T}. Such heating and mechanical mixing may significantly alter the disk's thermal and chemical structure, influencing grain growth, the distribution of volatiles, and the onset of planetesimal formation \citep{2017A&A...599A.113M,2025Natur.643..649M}. These feedback-driven changes could ultimately impact the initial conditions for planet formation and the chemical composition of future planetary systems. Further high-sensitivity observations and complementary disk modeling are essential to fully characterize the impact of outflow feedback on the protostellar disk, particularly during the critical early stages of planet formation.

\section{Summary}

In this study, we present a comparative analysis of infrared (\textit{JWST}/MIRI) and submillimeter (ALMA) emission from the jet/outflow system G208S3, also known as HOPS~315, located in the Orion molecular cloud. To facilitate this study, we developed a custom analysis tool \texttt{peakmoment} capable of consistently identifying bright spectral lines in \textit{JWST} MIRI datacubes and producing continuum-subtracted moment-0 maps. Our main findings are summarized below:

\begin{itemize}
\item \textit{JWST}/MIRI observations reveal two distinct temperature components in the outflow/jet, traced by H$_2$ rotational transitions. The excitation analysis indicates a mixture of warm (\(773 \pm 44~\mathrm{K}\)) and hot (\(2499 \pm 721~\mathrm{K}\)) gas. The visual extinction toward the emitting region is estimated to be \(A_V \sim 23.5 \pm 2.5\) mag. The combined molecular hydrogen column density is \(N(\mathrm{H}_2) = (3.48 \pm 1.02) \times 10^{20}~\mathrm{cm}^{-2}\). From the derived parameters, we estimate a jet mass-loss rate for the blueshifted lobe of \(\dot{M}_{\rm J, blue, JWST} \approx (0.27 \pm 0.20) \times 10^{-6}~M_\odot~\mathrm{yr}^{-1}\).
    
    
\item A combined analysis of the \textit{JWST}/MIRI and ALMA data reveals that the shorter-wavelength H$_2$ rotational lines (S(7) to S(4)) trace the collimated inner jet, spatially coincident with the high-velocity SiO emission detected by ALMA. In contrast, the longer-wavelength H$_2$ transitions (S(3) to S(1)) primarily trace the wide-angle outflow cavity, similar to the low-velocity CO emission observed with ALMA. Notably, ALMA detects a bipolar jet structure in both SiO and CO, while \textit{JWST}/MIRI reveals only the blueshifted component, likely due to extinction effects obscuring the redshifted lobe in the infrared. 

\item By comparing the spatial structure of the emission ``knots" in ALMA and \textit{JWST}/MIRI observations, we find that the infrared H$_2$ lines most likely trace the forward layer of the shock front, where the gas is hotter and more excited. In contrast, ALMA SiO and CO emission appear to trace regions just behind this forward shock layer, corresponding to cooler and denser post-shock gas. 
 

\item The protostellar system G205S3 is classified as an evolved Class~I source based on its spectral energy distribution (SED). However, several jet-related characteristics suggest that it may be at a younger evolutionary phase than implied by its SED. In particular, the detection of a well-collimated SiO jet with ALMA and the moderately high mass-loss rate estimated independently from both \textit{JWST}/MIRI and ALMA CO observations are more typical of Class~0 protostars. This apparent discrepancy may be explained by the system's geometry: the inclination and orientation may expose the central protostar more directly, enhancing the observed infrared emission and thereby mimicking the SED of a more evolved Class~I object.

\end{itemize}

\begin{acknowledgments}
I thank the anonymous referee for the constructive comments, which helped improve the clarity and robustness of this work. 
I would like to thank Dr. Chin-Fei Lee, Dr. Sheng-Yuan Liu, Dr. Tie Liu, and other members of the ALMASOP team for their support with ALMA data access, computational resources, and for valuable discussions regarding the data and scientific interpretation at various stages of my  research  over the past few years.
This work is based on observations made with the NASA/ESA/CSA James Webb Space Telescope. The data were obtained from the Mikulski Archive for Space Telescopes at the Space Telescope Science Institute, which is operated by the Association of Universities for Research in Astronomy, Inc., under NASA contract NAS 5-03127 for JWST. These observations are associated with JWST GO Cycle 1 (PI: Melissa McClure). The data presented in this paper were obtained from the Mikulski Archive for Space Telescopes (MAST) at the Space Telescope Science Institute. All the {\it JWST} data used in this paper can be found in MAST: \dataset[https://doi.org/10.17909/52vn-6191]{https://doi.org/10.17909/52vn-6191}. The codes utilized to analyse JWST IFU cubes are available in \dataset[https://doi.org/10.5281/zenodo.16623946]{https://doi.org/10.5281/zenodo.16623946} and \dataset[https://github.com/somastro/peakmoment]{https://github.com/somastro/peakmoment}. 
This paper makes use of the following ALMA data set: ADS/JAO.ALMA\# 2018.1.00302.S. (PI: Tie Liu). ALMA is a partnership of the ESO (representing its member states), the NSF (USA) and NINS (Japan), together with the NRC (Canada) and the NSC and ASIAA (Taiwan), in cooperation with the Republic of Chile. The Joint ALMA Observatory is operated by the ESO, the AUI/NRAO, and the NAOJ. The authors thank to the ALMA staff for their excellent support.

\end{acknowledgments}

\begin{contribution}
S.D. performed the archival data acquisition, data reduction, and analysis of the \textit{JWST} and ALMA observations. S.D. also conducted the interpretation of the results and wrote the manuscript.


\end{contribution}

%
\facilities{JWST, ALMA}

\software{astropy \citep{2013A&A...558A..33A,2018AJ....156..123A,2022ApJ...935..167A}, scipy \citep{2020NatMe..17..261V}, matplotlib \citep{2007CSE.....9...90H}, spectral-cube \citep{2016ascl.soft09017R}, reproject \citep{2018zndo...1162674R}, peakmoment\citep{2025zndo...16623946D}
}


\appendix
\renewcommand{\thefigure}{A\arabic{figure}}
\setcounter{figure}{0}

\section{Appendix information}\label{sec:Appendix_ALMA_cosiocont}
\begin{figure}
    \centering
    \includegraphics[width=1\linewidth]{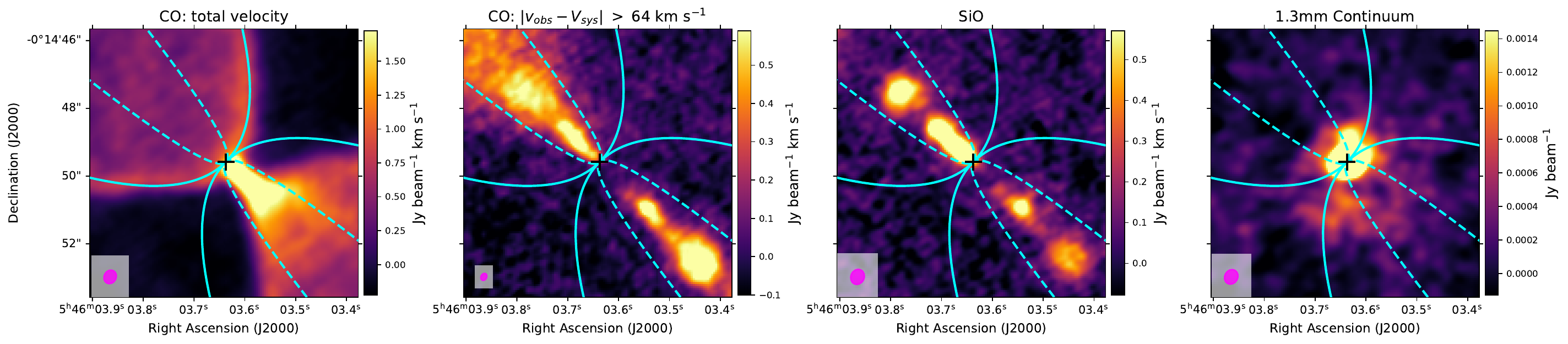}
    \caption{ALMA observations of the G205.46$-$14.56S3 region, zoomed in over an area of approximately $8\arcsec \times 8\arcsec$.
    Panel 1 shows the integrated CO emission tracing the full velocity extent of the molecular outflow. Panel 2 highlights the high-velocity $|V_{observed} - V_{systematic}| > 64 km s^{-1}$ CO emission associated with the jet-driven component.  Panel 3 presents the SiO moment map, tracing shocked gas consistent with recent outflow activity. Panel 4 displays the continuum emission, revealing compact dust structures including disk and envelope features. Crosshairs mark the protostellar position. Curved solid parabola trace the extended outflow cavities similar as Panel 1, while narrow dotted parabolas indicate high-velocity CO structures similar to panel 2. Synthesized beam sizes are shown as magenta ellipses in each panel for reference. Colorbars indicate the intensity units, as labeled individually. 
}
    \label{fig:Appendix_ALMA_cosiocont}
\end{figure}

Figure \ref{fig:Appendix_ALMA_cosiocont} shows ALMA observations of the G205.46$-$14.56S3 region in CO at different velocities (panels 1 and 2), SiO (panel 3), and 1.3,mm continuum. We observe that the CO emission as a whole delineates a wide-angle outflow (panel 1), while the high-velocity CO primarily traces a collimated axial jet. The morphology of the high-velocity CO emission closely resembles that of the SiO emission, indicating that both trace the collimated jet \citep[see also][]{2022ApJ...925...11D}.

\bibliography{sample701}{}
\bibliographystyle{aasjournalv7}



\end{document}